\documentclass[runningheads]{llncs}
\usepackage{graphicx}
\usepackage{comment}
\usepackage{amsmath,amssymb} % define this before the line numbering.
\usepackage{color}

\begin{document}
	% \renewcommand\thelinenumber{\color[rgb]{0.2,0.5,0.8}\normalfont\sffamily\scriptsize\arabic{linenumber}\color[rgb]{0,0,0}}
	% \renewcommand\makeLineNumber {\hss\thelinenumber\ \hspace{6mm} \rlap{\hskip\textwidth\ \hspace{6.5mm}\thelinenumber}}
	% \linenumbers
	\pagestyle{headings}
	\mainmatter
	\def\ECCVSubNumber{2857}  % Insert your submission number here
	
	\title{Multi-Dimension Fusion Network for Light Field Spatial Super-Resolution using Dynamic Filters} % Replace with your title

	% CAMERA READY SUBMISSION
	\titlerunning{Multi-Dimension Fusion Network for LFSSR using Dynamic Filters}
	% If the paper title is too long for the running head, you can set
	% an abbreviated paper title here
	%
	\author{Qingyan Sun \and 
	Shuo Zhang \and 	
	Song Chang \and 
	Lixi Zhu \and
	Youfang Lin}
	%
% 	\authorrunning{Lixi Zhu et al.}
	% First names are abbreviated in the running head.
	% If there are more than two authors, 'et al.' is used.
	%
	\institute{Beijing Jiaotong University\\
	\email{\{sunqingyanstc,zhangshuo,changsong,lixizhu,yflin\}@bjtu.edu.cn}}
	%******************
	\maketitle
	
	\begin{abstract}
		Light field cameras have been proved to be powerful tools for 3D reconstruction and virtual reality applications.
		However, the limited resolution of light field images brings a lot of difficulties for further information display and extraction.  
		In this paper, we introduce a novel learning-based framework to improve the spatial resolution of light fields.
		First, features from different dimensions are parallelly extracted and fused together in our multi-dimension fusion architecture. 
		These features are then used to generate dynamic filters, which extract sub-pixel information from micro-lens images and also implicitly consider the disparity information.
		Finally, more high-frequency details learned in the residual branch are added to the upsampled images and the final super-resolved light fields are obtained.
		Experimental results show that the proposed method uses fewer parameters but achieves better performances than other state-of-the-art methods in various kinds of datasets.
		Our reconstructed images also show sharp details and distinct lines in both sub-aperture images and epipolar plane images.
		% We use less than one-sixth of the parameters than the most advanced method and perform better in all test sets. We achieve $43.01$ dB of PSNR in $7 \times 7$ light field images for $\times 2$ super-resolution, which is the highest level in light field spatial super-resolution.
	\end{abstract}

	\section{Introduction}
	
	With the development of camera imaging technology, Light Field~(LF) cameras have been proved to be powerful tools for 3D reconstruction. 
	LF images record not only the intensity but also the direction of light in space. 
	Therefore, different sub-aperture images (SAIs) in LFs are available, which capture scenes in slightly different directions.  
	Such information has brought lots of new applications, such as image refocusing~\cite{ng2005light} and virtual reality applications~\cite{Lytro,Raytrix,plex_vr}.
	
	% Large-disparity LF cameras are used in many application scenarios, such as phones with multi-cameras.
	
	However, the development of optical lenses limits the imaging performance of LF cameras. 
	Usually, the spatial resolution of the LF camera is much lower than traditional cameras. 
	LR images not only affect visual perception but also bring a lot of difficulties for further information extraction, such as depth estimation. 
	In order to improve the spatial resolution of LF images, designing different methods for Light Field Spatial Super Resolution~(LFSSR) is very necessary.

	Since LF images record information from different dimensions, \textit{i.e.} angular and spatial, it is important to integrate all the information together for LFSSR.  
	However, how to explore and construct the relationship between multiple dimension data is still a difficult problem. 
	Some learning-based methods~\cite{farrugia2019light,cheng2019light} try to explicitly estimate disparity information between different SAIs and warp all SAIs to find corresponding relationships to supply sub-pixel details.
	Other methods~\cite{yeung2018light,zhang2019residual,wang2018lfnet,yoon2017light} propose end-to-end network structures to combine features from different dimensions together, in which disparity information is implicitly learned.
	However, in the above methods, the features from multiple dimensions in LFs cannot be fully extracted and adequately integrated. Moreover, all the above methods produce the high-resolution SAIs by mixing values from different SAIs together through a simple Convolutional Neural Networks (CNN), which ignores the disparity information and leads to a blurry output.

	\begin{figure}[!h]
		\centering
		\includegraphics[width=\textwidth]{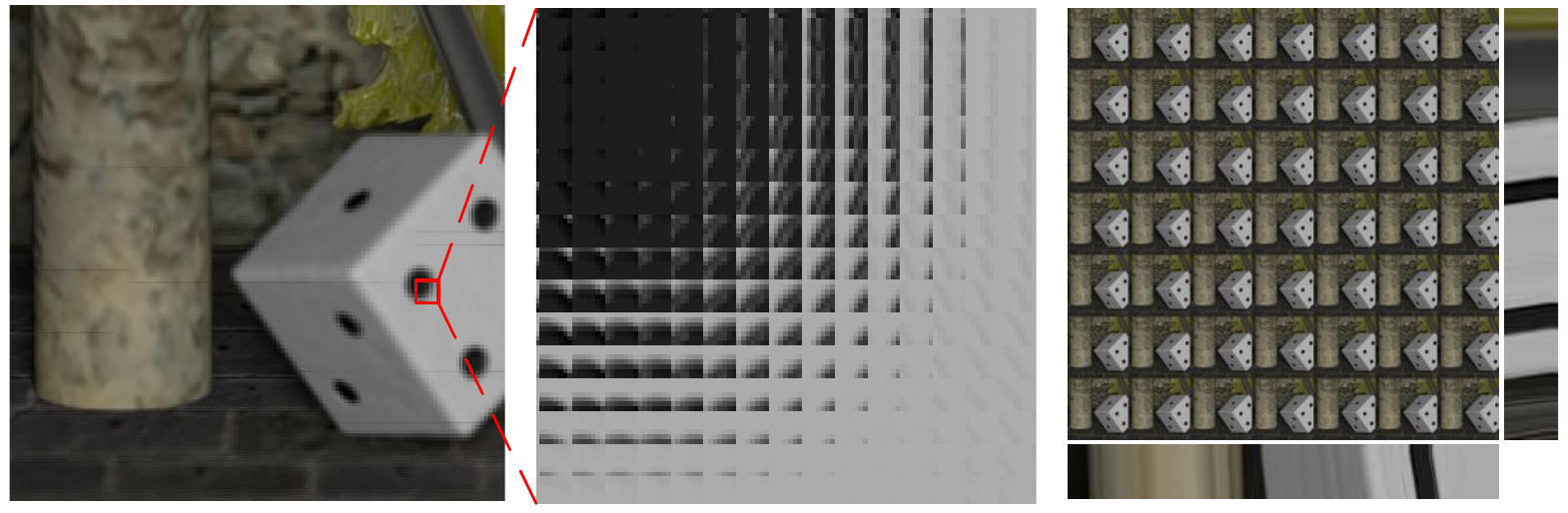}
		\begin{minipage}{.001\linewidth}
			\centerline{} 
		\end{minipage}
		\begin{minipage}{.32\linewidth}
			\centerline{(a) Micro-lens image} 
		\end{minipage}
		\begin{minipage}{.32\linewidth}
			\centerline{(b) Zoom in} 
		\end{minipage}
		\begin{minipage}{.32\linewidth}
			\centerline{(c) SAIs and EPIs} 
		\end{minipage}
		\caption{LF images in different dimensions. (a),(b) The micro-lens imags of LF images $I(:,:,x,y)$, including the angular information. The perspective minor changes in micro-lens help us rebuild missing details. (c) SAIs $I(u,v,:,:)$ and EPIs $I(u,:,x,:)$, $I(:,v,:,y)$. SAIs can be regarded as different views, which capture scenes from different directions. EPIs intuitively reflect the continuous changes of pixel in the same row (column).}
		\label{epi}
	\end{figure}
	
	In this paper, we proposed a novel end-to-end network, which can parallelly extract LF features from four different dimensions and adequately fuse them together for LFSSR.
	Instead of explicitly computing disparities between SAIs, the disparity information is implicitly learned to generate dynamic filters. The proposed dynamic filters are directly implemented on micro-lens images of the low-resolution LFs to recover sub-pixel details for high-resolution LFs.
	The residual information supplies high-frequency details to generate sharp and detailed LFs.

	Experimental results show that our framework outperforms the other state-of-the-art methods in different datasets and both numerical and visual comparisons. 
	The parallax information contained in LF images shows great significance for Super-Resolution (SR) work. 
	Since we do not directly combine values from multiple SAIs, the super-resolved LFs show much sharper details and consistent relationships in Epipolar Plane Images (EPIs).
	Specifically, the average PSNR of our framework on 57 LFs in \textit{General} is almost $1$ dB higher than the most advanced method, in which we only use less than one-sixth of their parameters.

	\section{Related Works}
	
	In this section, we briefly introduce state-of-the-art methods for LFSSR, including traditional and learning-based methods.
	
	Since SAIs in LFs capture scenes in slightly different directions, traditional LFSSR methods try to find sub-pixel information in SAIs based on disparity information. 
	Based on various disparity estimation methods~\cite{johannsen2017taxonomy}, some methods warped different SAIs in the reference SAI~\cite{bishop2009light,cho2013modeling,rossi2018geometry} and others interpolated lines in EPIs~\cite{wanner2012spatial,wanner2014variational}. The higher resolution SAIs are then obtained by constructing a global optimization problem and using complex optimization methods.
	Based on micro-lens images, Zhang \textit{et al.}~\cite{zhang2018micro} proposed to find sub-pixel from micro-lens images to rebuild missing information in SAIs. 
	However, since disparity information is needed for these methods, the reconstruction qualities are also easily affected by wrongly estimated disparities, especially around occluded regions. 
	% view may contain sub-pixels of SAI view,   
	% Though the framework is a traditional algorithm, finding sub-pixels from micro-lens is a worth trying method. 
	
	Since learning-based methods for a single image~\cite{lim2017enhanced} and video~\cite{jo2018deep} SR are rapidly developed and achieve good performances, other methods tried to design CNNs for LFSSR.
	Using optical flow or disparity estimation networks, some methods ~\cite{farrugia2019light,cheng2019light} proposed to warp SAIs in their networks and super-resolve each SAI using Very Deep Super-Resolution methods (VDSR)~\cite{kim2016accurate}. However, lots of pre-processes and post-processes are needed and make the networks too complex. 
	By contrast, some end-to-end networks are proposed for LFSSR with implicit disparity information. 
	Yeung \textit{et al.}~\cite{yeung2018light} proposed a Spatial-Angular Separable (SAS) convolution network. The network processes convolution alternately between the spatial plane and the angular plane. Then, a transposed convolutional layer upsamples the feature maps to a finer resolution. The SAS network is designed using fewer parameters but achieves similar performance compared with the similar networks using 4D convolutions.
	Wang \textit{et al.}~\cite{wang2018lfnet} developed a bidirectional recurrent network based on EPIs in horizontal and vertical directions. SAIs in the two EPIs are first super-resolved and other SAIs are super-resolved afterward.
	To make the most use of residual information, Zhang \textit{et al.}~\cite{zhang2019residual} proposed a Residual Network for Light Field (resLF). In resLF, each single SAI is super-resolved using different directions SAIs stack.  
	Each SAIs stack contains continuous changes in different directions of view. The resLF extracts features from each stack and merges them to extract all directions features. Then the global residual is added to reconstruct the SR center view image. 
	However, all the networks work on either spatial and angular domain\cite{yeung2018light}, or EPIs~\cite{wang2018lfnet} and cannot effectively integrate all kinds of features together. Moreover, since disparity information is unavailable, it is difficult for these networks to find specific features for pixels in different disparities.

	For video SR, lots of methods are proposed by considering the motion information between different frames. Similarly, some methods~\cite{kim2018spatio,caballero2017real,xue2019video} use optical flow to align each frame and then design different structures to fuse all features from different frames together. Other methods choose implicit motion compensation using dynamic filtering~\cite{jo2018deep} or deformable convolution~\cite{wang2019edvr}.
	Specifically, Jo \textit{et al.}~\cite{jo2018deep} proposed a method using dynamic filters. They proposed a weight shared network to extract features, then used features to calculate dynamic filters and residual information. Different from other methods that using bicubic to directly upsample each frame, the designed dynamic filters combine the moving information for upsample operation. Then the upsampled image and residual are summed up to obtain one super-resolved reference frame. 
	% ???????  Since the input of the network is SR frame and its neighborhoods, the output frames still contain the continuity of time.
	
	In this paper, we propose a new network to extract different kinds of features in multiple dimensions together and fuse them together for LFSSR.
	Moreover, since in LF images, SAIs have continuous view changes, which is similar for videos that have continuous motion change, we also especially design dynamic filters for LF images. The dynamic filters are implemented on micro-lens images to find spatial information from the angular domain. 
	
	\begin{figure}[!h]
		\centering
		\includegraphics[width=0.99\textwidth]{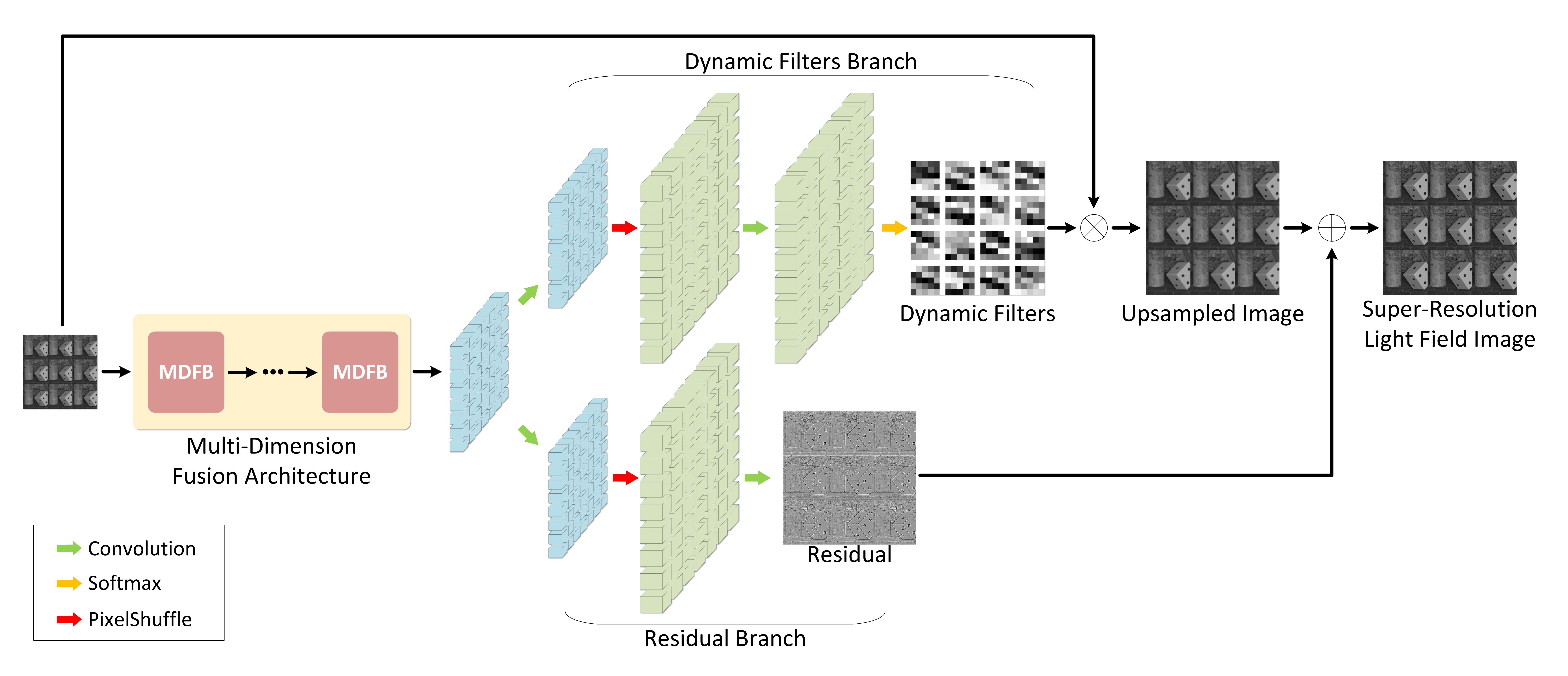}
		\caption{The proposed Multi-Dimension Fusion Network pipeline, which including Multi-Dimension Fusion Architexture, Dynamic Filter Branch and Residual Branch. Dynamic Filters $\mathcal{F}_d$ are generated to find spatial sub-pixels from micro-lens image and upsample the original LFs. Residual part $I^{R}$ is designed to add high-frequency details for the final recovered LFs.}
		\label{network}
	\end{figure}
	
	\section{Method}

	With 4 dimensions composition, LF images record both the angular and intensity information of light. LF images can be represented as $I(u,v,x,y)$ , where $(u,v)$ denotes the angular plane and $(x,y)$ denotes the spatial plane. Assuming that the resolution of $I$ is described with $(U,V,X,Y)$, \textit{i.e.} $I\in\mathbb{R}^{U\times V\times X\times Y}$, the goal of LFSSR is to reconstruct a Low Resolution (LR) image $I^{lr}\in\mathbb{R}^{U\times V\times X\times Y}$ into a SR image $I^{sr}\in\mathbb{R}^{U\times V\times rX\times rY}$, where $r$ represents the upsampling factor in SR. 
	% The low-resolution images $I^{sr}$ are downsampled from the corresponding ground truth high-resolution(HR) images $I^{hr}\in\mathbb{R}^{U\times V\times rX\times rY}$. 
	We convert images to YCbCr color space and only deal with Y channel images. With our network $G(\cdot)$, the SR image can be represented as:
	
	\begin{equation} 
	I^{sr}=G(I^{lr})
	\end{equation}

	\subsection{Network Overview}

	The proposed network is composed of Multi-Dimension Fusion Network (MDFN), Dynamic Filters Branch (DFB) and Residual Branch (RB), as shown in Fig.~\ref{network}. 
	The proposed Multi-Dimension Fusion Network $G^{M\!D\!F\!N}(\cdot)$ is designed concisely and effectively to learn sub-pixel features in parallel from $4$ different domains. 
	MDFN $G^{M\!D\!F\!N}(\cdot)$ captures full-light-field features $F_{n}$ from LR image $I^{lr}$, where $n$ represents the number of depth of blocks. 
	\begin{equation} 
	F_{n}=G^{M\!D\!F\!A}(I^{lr})
	\end{equation}

	Based on $F_{n}$, DFB $G^{D\!F\!B}(\cdot)$ is designed to generate dynamic filters $\mathcal{F}_d$ for each pixel in SR image $I^{sr}$, which consider the specific disparity information of each pixel.
	The generated dynamic filters are then implemented on the micro-lens images of original LR images to calculate the upsampled image $I^{U}$. 
	RB $G^{R\!B}(\cdot)$ learns high-frequency residual information $I^{R}$ through LF features $F_{n}$. 
	The output SR image $I^{sr}$ is the sum of the residual image $I^{R}$ and the dynamic filters upsampled image $I^{U}$: 
	% The whole network pipeline can represented as:
	\begin{equation} 
	I^{sr}=I^{U}+I^{R}=G^{D\!F\!B}(F_{n})\bigotimes I^{lr}+G^{R\!B}(F_{n})
	\label{IUIR}
	\end{equation}

	\begin{figure}[!h]
		\centering
		\includegraphics[width=\textwidth]{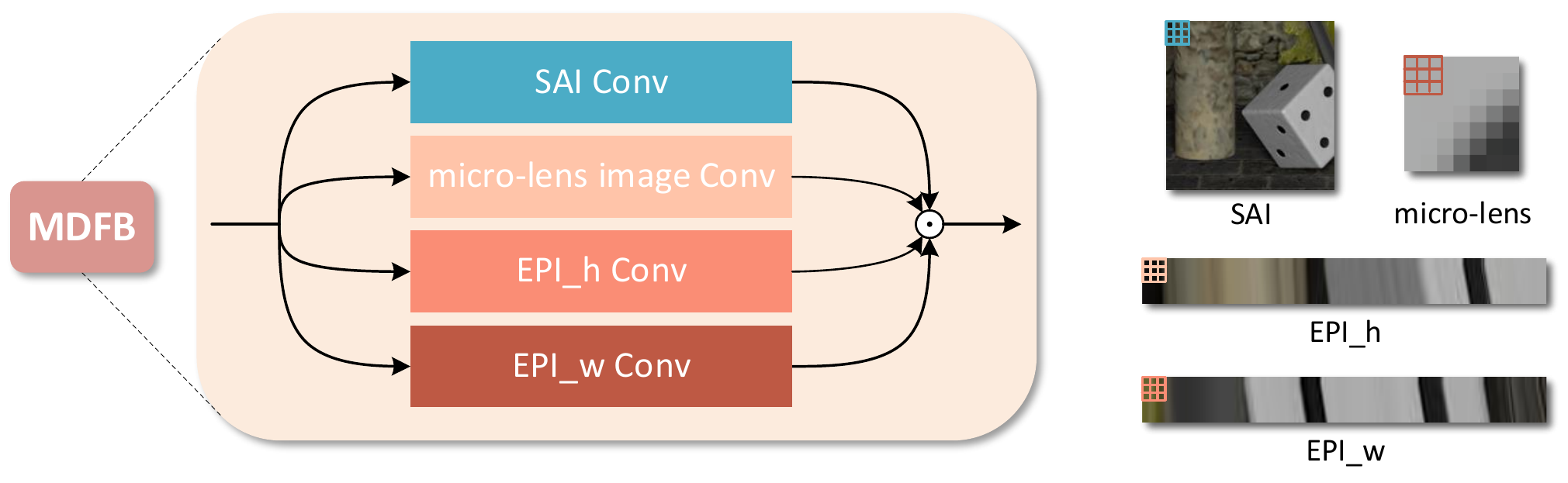}
		\caption{The proposed Multi-Dimension Fusion Block. $4$ convolutions are designed to extract features in different dimensions, \textit{i.e.} SAIs, micro-lens imags, horizontal and vertical EPIs, in parallel. All features are then concatenated for the next block. }
		\label{MDFB}
	\end{figure}

	\subsection{Multi-Dimension Fusion Architecture}

	Since 4D LF records scene information both in spatial and angular domain, how to effectively investigate sub-pixel information from different domains is important for LFSSR.
	% There are 4 dimensions in LF images, while the traditional Convolution can only process 3D input at most. 
	To process the 4D input, the most intuitive way is to construct 4D convolutions, as in~\cite{yeung2018light}. 
	However, the plain 4D convolution requires a huge amount of calculations and parameters. At the same time, without prior, the network converges slowly and performs poorly when processing the huge input. 
	Other networks are constructed elaborately by exploring features either in angular and spatial alternately~\cite{yeung2018light} or in EPIs from different directions seperately~\cite{wang2018lfnet}. 
	However, these complex networks cannot learn implicit information with too many prior conditions and ignore necessary information from other domains.
	In this paper, we design a new structure for LF images, which explore sub-pixel features from different domains in parallel.
	
	% However, the complex network also brought so much priori that some of the implicit information may not be learned by the network. 
	% To design a delicate network, keeping original structural information and adding some necessary priors is helpful.
	
	The proposed MDFN is made up of Multi-Dimension Fusion Block (MDFB), in which 4 different 2D convolution layers $G_i^{S}(\cdot ),G_i^{A}(\cdot ),G_i^{E_{h}}(\cdot ),G_i^{E_{v}}(\cdot )$ are used. MDFB extracts features from these convolution layers parallelly. 
	The input of the next block is the output of the previous block. 
	For more convenient representation, we consider the input of the network $I^{lr}(u,v,x,y)$ as the 0-th feature $F_{0}(u,v,x,y)$.
	
	% ​    The 4 dimensions structure of LF images contains much information. 
	If the angular plane is fixed, \textit{i.e.} $u=u^{*},v=v^{*}$, in LF images, the image $I(u^{*},v^{*},:,:)$ is converted into multiple SAIs. Similar with other ordinary images, SAIs contain spatial information of scenes.
	The spatial convolution $G_i^{S}(\cdot )$ is designed to extract spatial features in each SAI:
	\begin{equation} 
	F^{S}_{i}(u^{*},v^{*},:,:)=G_i^{S}(F_{i-1}(u^{*},v^{*},:,:)).
	\end{equation}

	If the spatial plane is fixed, \textit{i.e.} $x=x^{*},y=y^{*}$, the micro-lens images $I(:,:,x^{*},y^{*})$ are obtained. In LF images, pixels in micro-lens images come from the same position in each SAIs and contain angular information. Moreover, micro-lens images are corresponding to different regions in SAIs based on their disparities as in~\cite{zhang2018micro}.
	The angular convolution $G_i^{A}(\cdot )$ is designed to extract angular features in each micro-lens images:
	\begin{equation} 
	F^{A}_{i}(:,:,x^{*},y^{*})=G_i^{A}(F_{i-1}(:,:,x^{*},y^{*})).
	\end{equation} 
	
	If one dimension of spatial and the corresponding angular plane is fixed, \textit{i.e.} $u=u^{*},x=x^{*}$ or $v=v^{*},y=y^{*}$, then EPIs $I(u^{*},:,x^{*},:)$ and $I(:,v^{*},:,y^{*})$ are obtained. 
	In EPIs, pixels from angular and spatial domains are considered simultaneously and sub-pixel information in one specific direction is easy to find in other views~\cite{zhang2019residual}, which also intuitively reflect the disparity informaion~\cite{zhang2016robust}. 
	The EPI convolutions $G_i^{E_{h}}(\cdot ), G_i^{E_{v}}(\cdot )$ are used to extract structural features in each EPI:
	\begin{equation} 
	F^{E_{h}}_{i}(u^{*},:,x^{*},:)=G_i^{E_{h}}(F_{i-1}(u^{*},:,x^{*},:)).
	\end{equation} 
	\begin{equation} 
	F^{E_{v}}_{i}(:,v^{*},:,y^{*})=G_i^{E_{v}}(F_{i-1}(:,v^{*},:,y^{*})).
	\end{equation} 
	
	When $4$ different convolutions are implemented, we reshape all the features into the same size and concatenate them as the input for the next MDFB:
	\begin{equation} 
	F_{i}=[F^{S}_{i},F^{A}_{i},F^{E_{h}}_{i},F^{E_{v}}_{i}]
	\end{equation}
	
	Since the MDFB extracts features from different dimensions and all these features are adequately integrated afterward, the designed structure is able to fully extract the needed sub-pixel information for the SR. 
	After $n$ MDFBs, the multi-dimention features $F_{n}\in\mathbb{R}^{ U\times V\times X\times Y}$ are obtained.

	\begin{figure}[!h]
		\centering
		\includegraphics[width=\textwidth]{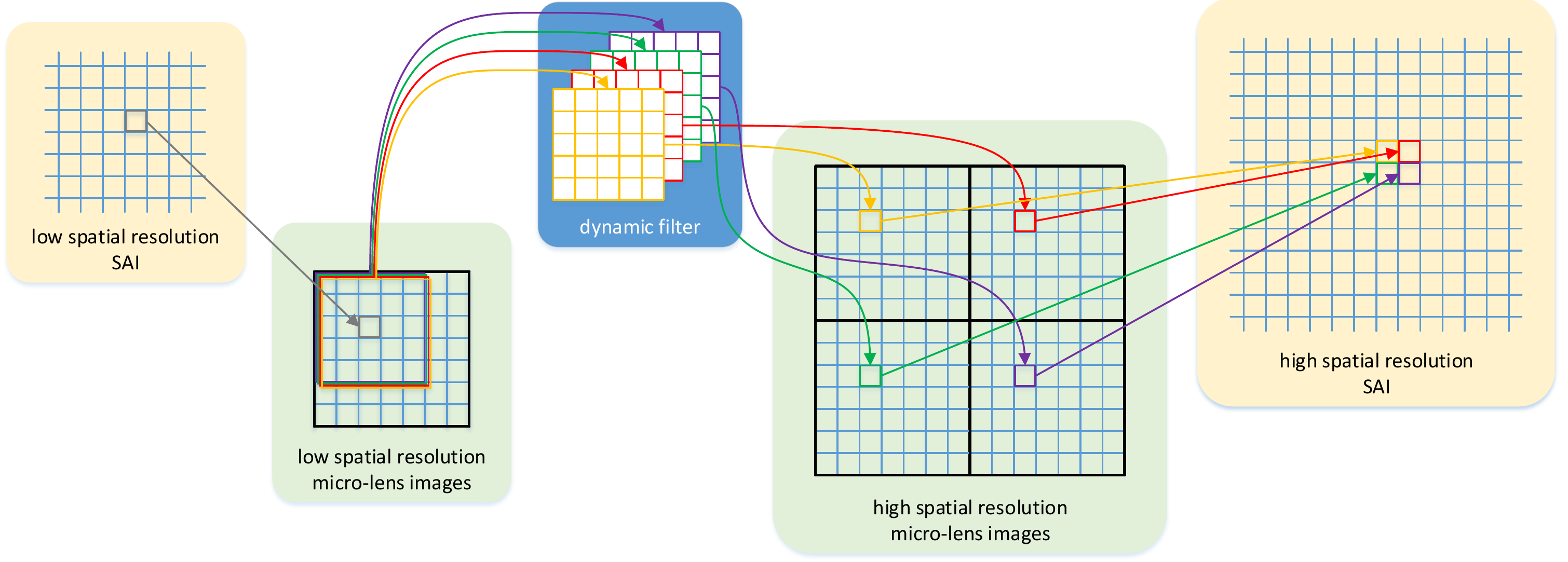}    
		\caption{The proposed dynamic upsampling filters. The figure shows one LF with $7\times 7$ angular resolution for $\times2$ SR. For each pixel in one SAI, $4$ dynamic filters are generated and implemented on the corresponding micro-lens image and produce spatial sub-pixel information for the corresponding upsampled SAI.}
		\label{dfconv}
	\end{figure}
	
	\subsection{Dynamic Filters Branch}
	
	The basic SR methods increase the image resolution by applying the bilinear or bicubic interpolation method. In CNN, the deconvolution filters~\cite{yeung2018light} and pixel-shuffle layers~\cite{zhang2019residual} are often used to upsample LR images into the desired resolution.
	However, since all pixels in one image are upsampled using the same filters, the specific features of each pixel are ignored.
	In LFSSR, as different SAIs are available, previous methods choose to upsample each SAI individually also using deconvolution filters~\cite{yeung2018light} or pixel-shuffle layers~\cite{zhang2019residual}.
	However, the relations between each SAIs are ignored and each image is upsampled only based on spatial information. 
	% and perform poorly in large scale super-resolution. 
	% The basically fixed upsampling filters limit their upsampling ability, as they cannot restore sharp and textured regions.
	In this paper, we propose to use the dynamic filters to upsample original LR input $I^{lr}$. 
	The dynamic filters upsample every pixel respectively by considering the relations between each SAIs so that dynamic filters can adaptively upsample LR images to generate sharp details images.

	In order to consider the disparity between each SAI, we design dynamic filters based on the multi-dimension feature $F_{n}$. 
	% Since upsample on all 4 dimensions makes the network much more complex and huge, and growth on parameters is unacceptable. 
	The dynamic filters are designed as 2D filters, which are implemented on each micro-lens image and provide sub-pixel information for each specific pixel in SAIs. 
	Inspired by~\cite{zhang2018micro}, we generate dynamic filters to find sub-pixels from micro-lens images to reconstruct sharp details SR images.
	As in Fig.~\ref{epi}, details in micro-lens images provide various sub-pixel information for SAIs according to the different disparities. 
	Different from traditional layers using the same filter for all pixels, the dynamic filters consider disparity and combine more information from micro-lens images.

	The DFB includes a 2D convolution layer, a pixel-shuffle layer, a 2D convolution layer and a softmax layer in order. All the 2D convolutions are implemented on SAIs. The softmax layer is used for normalization and make the model converge faster.
	For $\times r$ super-resolution tasks, $U\times V\times rX\times rY$ dynamic filters are generated for each pixel in one light field and each filter size is $d\times d$, where $d$ is kernel size of each dynamic filter. 
	Finally, each output pixel is created by local filtering on an LR pixel in the input LR image $I^{lr}$ with the corresponding filter as follows:
	
	\begin{align}
	&I^{U}(u,v,rx+\Delta x,ry+\Delta y) =  \notag \\
	&\sum_{i=1}^{d}\sum_{j=1}^{d}{ \mathcal{F}_d (u,v,rx+\Delta x,ry+\Delta y,i,j)\cdot I^{lr}(u+i-\left \lfloor \frac{d}{2}\right \rfloor,v+j-\left \lfloor \frac{d}{2}\right \rfloor,x,y)},
	\end{align}where $\Delta x,\Delta y\in[0,r)$ are coordinates in each $r\times r$ output block. The upsample operation only contains addition and multiplication in fixed positions, so it allows back-propagation.

	% Different from traditional methods, we neither need to perform depth estimation nor need depth information for supervision.
	
	\begin{figure}[!h]
		\includegraphics[width=\textwidth]{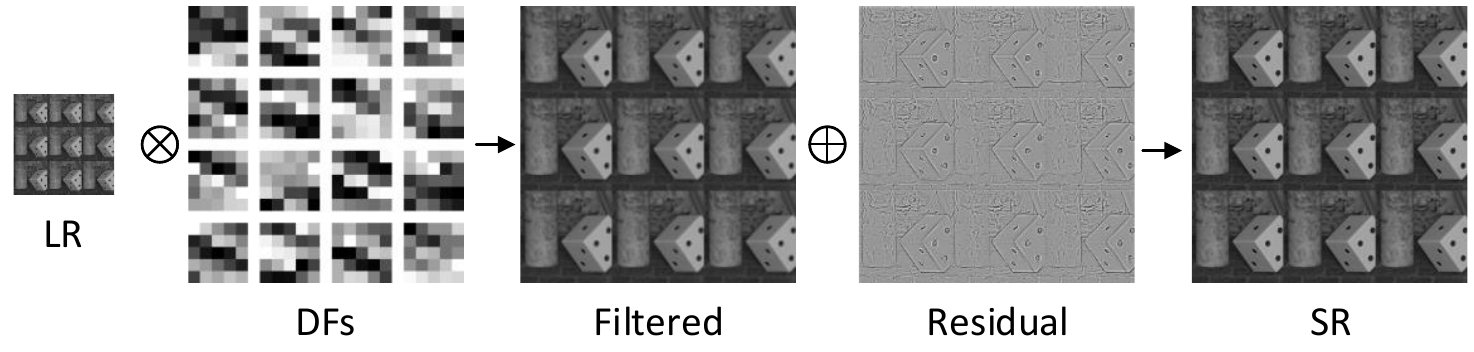}
		\caption{One examples of the upsampling process in our framework. $16$ dynamic filters are generated for $\times 4$ task to upsample the SAIs. The residual image $I^R$ is normalized for better visualization. High-frequency details $I^R$ are added to the upsampled SAIs $I^U$ and the final super-resolved LF $I^{sr}$ is obtained.}
		\label{residual}
	\end{figure}

	\subsection{Residual Branch}

	Since $I^{U}$ only contain sub-pixel information extracted from micro-lens images, some high-frequency information is still missing, especially for micro-lens images at the focus plane. 
	Moreover, the linear calculation is unable to provide sufficient texture details.
	Therefore, we also design the Residual Branch to further supplement residual information.
	RB is employed to estimate a residual image to increase non-linear high-frequency details.
	The residual information is also generated based on the multi-dimension features $F_{n}$. 
	The structure of RB includes 2 2D convolution layers on SAIs to decrease the number of channels, a pixel-shuffle layer for upsampling.
	RB is able to extract information between different dimensions, which guarantees consistency across dimensions.
	As shown in Fig.~\ref{residual}, the residual image $I^{R}$ contains high-frequency details to help reconstruct high contrast SR image $I^{sr}$.

	\section{Implementation}

	The designed MDFN has $n=8$ MDFBs and each MDFB has the same number of input and output channels $c=80$, except for $F_0 = I^{lr}$. The size of dynamic filters is $5\times 5$.
	The parallel convolutions in MDFBs $G^{S}(\cdot ),G^{A}(\cdot ),G^{E_{h}}(\cdot ),G^{E_{v}}(\cdot )$ have the same input and output channels, which is a quarter of the total number of channels ${c}/{4}=20$ . 
	The 3D convolutions in DFB and RB have the kernel size $(1,1,1)$.
	% , which concatenate at the angular plane. 
	All activation functions in MDFN are PReLU and we pad zero in all convolution layers.  
	The number of MDFN parameters is slightly affected by the size of dynamic filters and the SR scale.
	We use L1 loss as loss function:
	\begin{equation} 
	\mathcal{L}^{L1}(P)=\frac{1}{N}\sum_{p\in P}|x(p)-y(p)|
	\end{equation} 
	
	To fairly compare with state-of-the-art methods, we use 130 LF dataset~\cite{yeung2018light} for training. 
	% which is the most used dataset in LFSR. 
	To make full use of the dataset, LF images are rotated in $90$, $180$, $270$ degree and are flipped horizontally and vertically randomly.
	In each training batch, the crop size in each SAI is $24\times 24$ and the batch size is $22$.
	The model is trained with Adam optimizer and all weights are initialized using Kaiming~\cite{he2015delving} method.
	The network is implemented using PyTorch. 
	Our model is trained roughly $3$ days using a GTX 2080Ti GPU.
	
	The source code and trained model will be public on GitHub soon.

	\section{Experimental Results}
	
	In the experiments, we test our models on different datasets, including real-world LFs from \textit{General Category}~\cite{Stanford2016}, \textit{EPFL}~\cite{rerabek2016new} and synthetic images from \textit{HCI1}~\cite{wanner2013datasets}, \textit{HCI2}~\cite{honauer2016benchmark}. 
	All LF images are first cropped with $7\times 7$ angular resolution and downsampled with $r=2$ or $4$ magnification factors using the protocol in~\cite{rossi2018geometry,yeung2018light} and then super-resolved to the higher resolution.
	The super-resolved LFs are compared with the original high-resolution LFs and 
	the average PSNR and SSIM~\cite{wang2004image} on Y channel overall SAIs are used for evaluation.
	
	In this section, we first visualize the dynamic filter in detail.
	Then the performances of multi-dimension fusion networks and DFs are further analyzed by designed different ablation experiments.
	The final results are compared with state-of-the-art methods, including the single image SR method EDSR~\cite{lim2017enhanced}, LFSSR methods LFNet~\cite{wang2018lfnet}, SAS~\cite{yeung2018light}, 4D~\cite{yeung2018light} and resLF~\cite{zhang2019residual} method. 
	
	\begin{figure}[!h]
		\includegraphics[width=\textwidth]{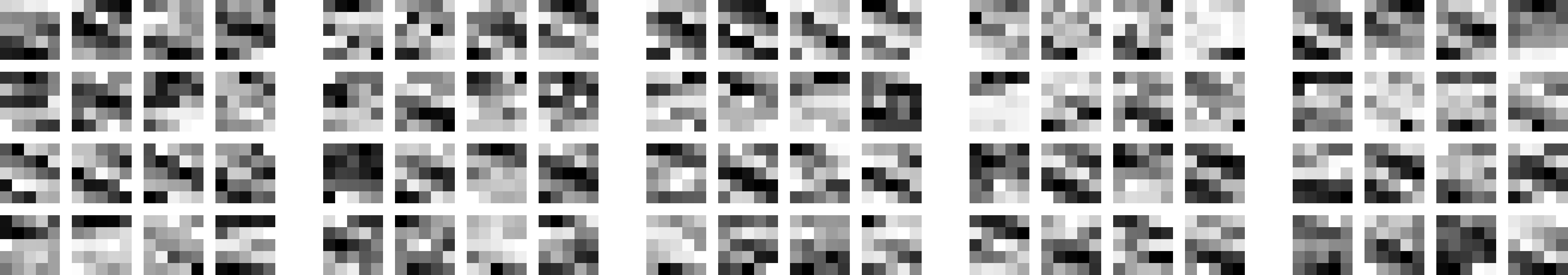}
		\caption{The dynamic filters for $\times 4$ task and each group includes $16$ filters of different pixels. The differences between each filter in one group show that the dynamic filters find different sub-pixel values from the same micro-lens image, so that the details of SR images are rich and sharp.}
		\label{allfilters}
	\end{figure}

	\subsection{Visualization of MDFN}

	Dynamic filters are designed to find sub-pixels information from micro-lens images. 
	Different from the deconvolution or pixel-shuffle layer, every pixel in each SAI has its own filter. 
	The upsampled sub-pixels are calculated by applying one specific filter on the corresponding micro-lens image. 
	As shown in Fig.~\ref{epi}, for pixels in different disparities, their micro-lens images corresponding to different regions in SAIs. 
	As analyzed in~\cite{zhang2018micro}, the sub-pixels information in micro-lens images can be used for improving resolution for SAIs if the disparities are estimated.
	Since the dynamic filters are obtained using CNNs, the rotation direction and occlusion information in the micro-lens image are not required anymore.
	In our DFB, the filters are dynamically learned to combine the disparity information implicitly and provide different values for around sub-pixels.
	As shown in Fig.~\ref{allfilters}, for $\times 4$ tasks, the $16$ filters are provided for one pixel in $I^{lr}$ and produce the sub-pixel values for the super-resolved $I^{sr}$.
	Different weights in $16$ filters show that the filters consider the corresponding relationships between SAIs and micro-lens images and provide different information for different sub-pixel points.
	Extracting sub-pixels form micro-lens images for SAIs provides another way to fuse information from different dimensions for LFSSR to reconstruct sharp and high contrast pixels.
	
	\subsection{Ablation Experiment}
	
	In this section, 57 LFs from \textit{General} dataset are used for comparison. 
	We first evaluate the performance multi-dimension fusion architacture by designing different kinds of features and extracting them in serial or parallel.
	Then, the performances of the propose dynamic filters are verified by comparing different upsampling methods.
	
	\begin{table}[!h]
		\centering
		\caption{Average PSNR/SSIM for $\times 2$ SR using features from different dimensions.}
		\begin{tabular}{c|c|c|c|c}
			\hline
			Methods & SAS~\cite{yeung2018light}  & Ours-SA & Ours-EPI & Ours\\
			\hline
			PSNR/SSIM & 41.87/0.982 & 42.76/0.985  &42.85/0.985 & \textbf{43.01}/\textbf{0.986}\\
			\hline
		\end{tabular}
		\label{ablationfusion}
	\end{table}
	
	\subsubsection{Multi-Dimension Fusion}

	In our MDFA, features from $4$ different dimensions of LFs are learned in parallel and fused together in each block. 
	% learns full-light-field features in the proposed network to keep the original structure. 
	% $4$ different plane convolutions propose all features at the same time and concatenate features to fusion information. 
	Feature parallelization and fusion are considered to be the key point that makes MDFA work.
	In this section, we use different features and fusion methods to verify our MDFA.
	
	We first design similar models using features from different dimensions, \textit{i.e.} spatail and angular features on micro-lens images $G_i^{A}$ and SAIs $G_i^{S}$ (labelled as Ours-SA) and EPI features on $2$ EPIs $G_i^{E_h},G_i^{E_v}$ (labelled as Ours-EPI). 
	For a fair comparison, Ours-SA, Ours-EPI is designed to have the same number of parameters with our original network. The related results are shown in Table.~\ref{ablationfusion}. 
	As analyzed, SAIs contain scenes texture and micro-lens images shows angular of light. %which including disparity information.
	By contrast, EPIs show continuous perspective changes and are able to directly reflect disparity information. 
	These images offer different information for LFSSR to extract full-light-field features. 
	Numerical results indicate that combining all features from $4$ dimensions together is better than just using features from $2$ dimensions. 
	% Convolutions on fewer planes will leads to a lack of information. 

	We then compare our parallel fusion method with the serial fusion method. 
	In~\cite{yeung2018light}, Yeung \textit{et al. } designed a SAS convolution to extract features from spatial and angular domain iteratively using a serial mode.
	As in Table.~\ref{ablationfusion}, compared SAS with Ours-SA, using features from the same dimensions, the parallel fusion method achieves better performances than SAS.
	%With more parameters, SAS still performs worse than ours.
	% Serial execution means fewer convolutions on the same plane than parallel execution. 
	By repeatly changing dimesions in spatial and angular domain when extracting features, the network cannot continuously learn deeper information in each dimension. 
	In contrast, in our model, features are simultaneously extracted in all dimensions and fused together in each block. The repeated blocks are able to fully explore information from different dimensions for LFSSR. 
	Parallelization is a strong strategy to process LF data. Though SAS has more parameters than Ours, it still performs worse than any of our parallel networks.
	% Fusion connects information from different dimensions. 

	\begin{table}[!h]
		\centering
		\caption{Average PSNR/SSIM for $\times 4$ SR using different upsampling methods.}
		
		\begin{tabular}{c|c|c}
			\hline
			Methods & Deconvolution  & Dynamic Filter\\
			\hline
			PSNR/SSIM & 35.95/0.940 & \textbf{36.01}/\textbf{0.940}\\
			\hline
		\end{tabular}
		\label{ablationdfs}
	\end{table}

	\subsubsection{Dynamic Filters}
	
	In order to verify that the proposed dynamic filter exploits the angular information based on disparities, we compare different upsampling methods in the same network.
	In~\cite{yeung2018light,zhang2019residual}, transposed convolutions are used to upsampled LR input.
	These methods upsample each SAI separately, which doesn't fuse angular information. 
	As in Table.~\ref{ablationdfs}, using the proposed dynamic filters for upsampling, the model achieves better performances than using the deconvolution upsampling method (labeled as Deconvolution).
	By extracting features from the micro-lens image, MDFN is able to obtain sub-pixels from different SAIs~\cite{zhang2018micro} and provide more details for high-resolution LF.

	\begin{table}
		\centering
		\caption{Average PSNR/SSIM in different datasets for $\times 2$ LFSSR.}
		\begin{tabular}{c|c|c|c|c|c|c}
			\hline
			\multicolumn{1}{r|}{} & LFNet~\cite{wang2018lfnet} & EDSR~\cite{lim2017enhanced} & SAS~\cite{yeung2018light} & 4D~\cite{yeung2018light} & resLF~\cite{zhang2019residual} & Ours \\
			\hline
			\hline
			\textit{HCI1} & 36.46/0.965 & 37.10/0.954 & 41.31/0.977 & 41.58/0.978 & 41.02/0.975 & \textbf{42.31/0.982} \\
			\hline
			\textit{HCI2} & 33.63/0.932 & 32.72/0.918 & 36.38/0.953 & 36.74/0.956 & 36.40/0.957 & \textbf{37.33/0.964} \\
			\hline
			\textit{EPEL} & 32.70/0.935 & 32.29/0.925 & 35.21/0.951 & 35.46/0.954 & 34.40/0.948 & \textbf{35.96/0.963} \\
			\hline
			\textit{General} & 36.56/0.953 & 37.22/0.961 & 41.87/0.982 & 42.09/0.982 & 40.77/0.978 & \textbf{43.01/0.986} \\
			\hline
		\end{tabular}%
		\label{x2compare}%
	\end{table}% 

	\subsection{Comparisons with Other Methods}

	\begin{figure*}[!t]
		\centering
		\scriptsize
		\begin{minipage}{.155\linewidth}
			\centerline{}
		\end{minipage}
		\begin{minipage}{.13\linewidth}
			\centerline{GT}
		\end{minipage}
		\begin{minipage}{.13\linewidth}
			\centerline{EDSR} 
		\end{minipage}
		\begin{minipage}{.13\linewidth}
			\centerline{resLF}
		\end{minipage}
		\begin{minipage}{.13\linewidth}
			\centerline{SAS}
		\end{minipage}
		\begin{minipage}{.13\linewidth}
			\centerline{4D}
		\end{minipage}
		\begin{minipage}{.13\linewidth}
			\centerline{Ours}
		\end{minipage}
		\begin{minipage}{.99\linewidth}
			\centerline{\includegraphics[width=\linewidth]{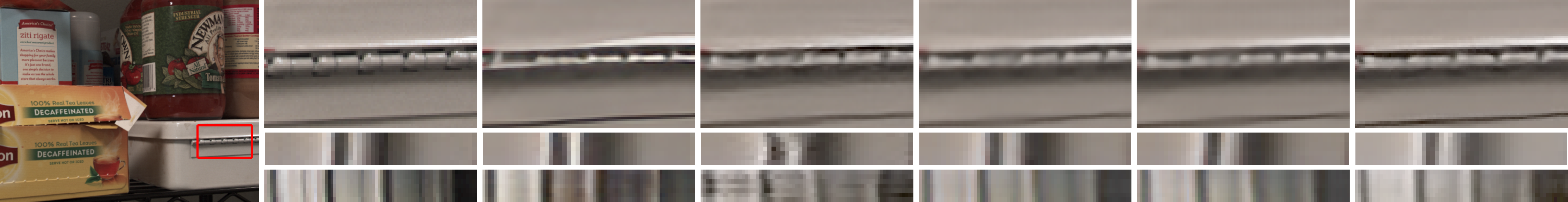}}
		\end{minipage}
		\begin{minipage}{.155\linewidth}
			\centerline{\textit{General 37}}
		\end{minipage}
		\begin{minipage}{.13\linewidth}
			\centerline{PSNR/SSIM}
		\end{minipage}
		\begin{minipage}{.13\linewidth}
			\centerline{$29.51$/$0.885$}
		\end{minipage}
		\begin{minipage}{.13\linewidth}
			\centerline{$31.45$/$0.912$}
		\end{minipage}
		\begin{minipage}{.13\linewidth}
			\centerline{$32.49$/$0.921$}
		\end{minipage}
		\begin{minipage}{.13\linewidth}
			\centerline{$32.65$/$0.926$}
		\end{minipage}
		\begin{minipage}{.13\linewidth}
			\centerline{$33.87$/$0.945$}
		\end{minipage}

		\begin{minipage}{.99\linewidth}
			\centerline{\includegraphics[width=\linewidth]{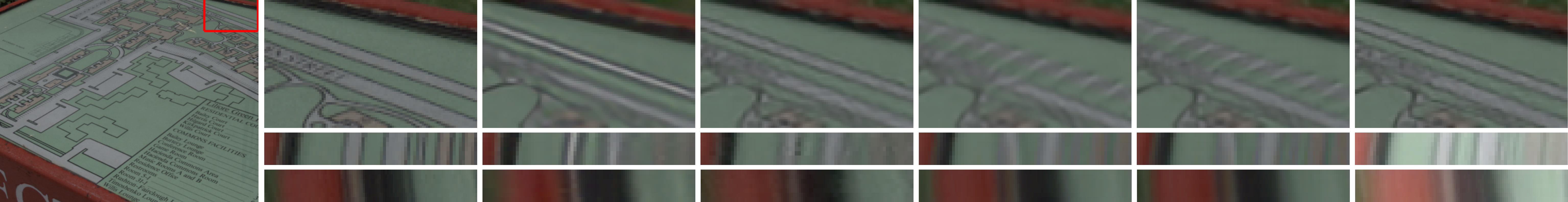}}
		\end{minipage}
		\begin{minipage}{.155\linewidth}
			\centerline{\textit{General 41}}
		\end{minipage}
		\begin{minipage}{.13\linewidth}
			\centerline{PSNR/SSIM}
		\end{minipage}
		\begin{minipage}{.13\linewidth}
			\centerline{$30.78$/$0.828$}
		\end{minipage}
		\begin{minipage}{.13\linewidth}
			\centerline{$31.98$/$0.839$}
		\end{minipage}
		\begin{minipage}{.13\linewidth}
			\centerline{$32.60$/$0.850$}
		\end{minipage}
		\begin{minipage}{.13\linewidth}
			\centerline{$33.02$/$0.871$}
		\end{minipage}
		\begin{minipage}{.13\linewidth}
			\centerline{$33.78$/$0.899$}
		\end{minipage}

		\begin{minipage}{.99\linewidth}
			\centerline{\includegraphics[width=\linewidth]{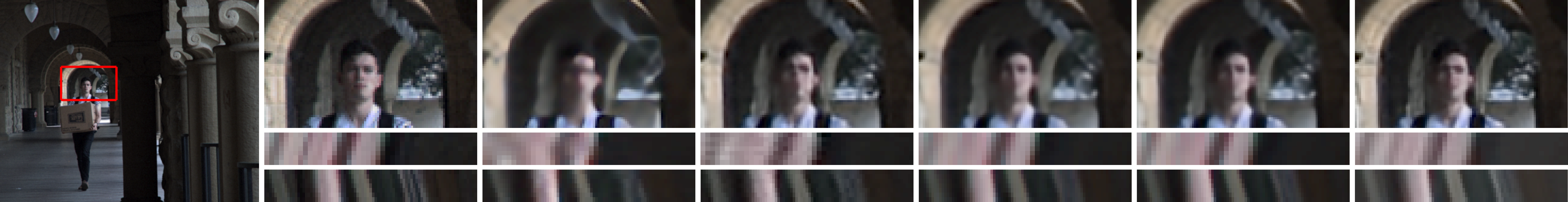}}
		\end{minipage}
		\begin{minipage}{.155\linewidth}
			\centerline{\textit{General 52}}
		\end{minipage}
		\begin{minipage}{.13\linewidth}
			\centerline{PSNR/SSIM}
		\end{minipage}
		\begin{minipage}{.13\linewidth}
			\centerline{$34.04$/$0.930$}
		\end{minipage}
		\begin{minipage}{.13\linewidth}
			\centerline{$37.19$/$0.950$}
		\end{minipage}
		\begin{minipage}{.13\linewidth}
			\centerline{$38.52$/$0.957$}
		\end{minipage}
		\begin{minipage}{.13\linewidth}
			\centerline{$38.60$/$0.960$}
		\end{minipage}
		\begin{minipage}{.13\linewidth}
			\centerline{$39.06$/$0.962$}
		\end{minipage}

		\begin{minipage}{.99\linewidth}
			\centerline{\includegraphics[width=\linewidth]{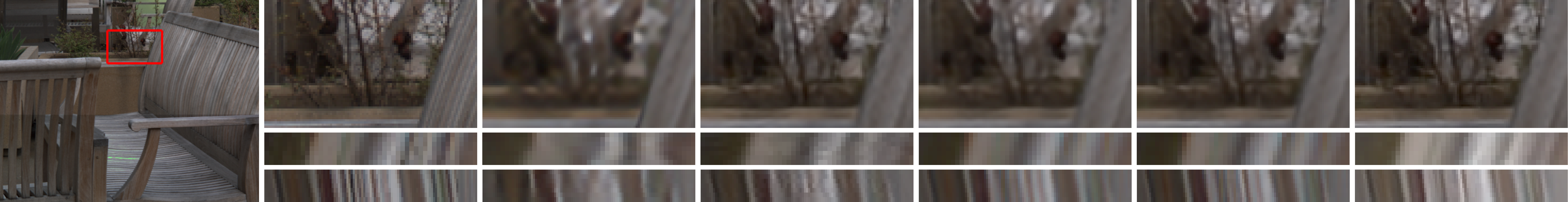}}
		\end{minipage}
		\begin{minipage}{.155\linewidth}
			\centerline{\textit{General 55}}
		\end{minipage}
		\begin{minipage}{.13\linewidth}
			\centerline{PSNR/SSIM}
		\end{minipage}
		\begin{minipage}{.13\linewidth}
			\centerline{$32.24$/$0.843$}
		\end{minipage}
		\begin{minipage}{.13\linewidth}
			\centerline{$33.22$/$0.860$}
		\end{minipage}
		\begin{minipage}{.13\linewidth}
			\centerline{$33.30$/$0.860$}
		\end{minipage}
		\begin{minipage}{.13\linewidth}
			\centerline{$33.54$/$0.874$}
		\end{minipage}
		\begin{minipage}{.13\linewidth}
			\centerline{$33.77$/$0.884$}
		\end{minipage}

		\begin{minipage}{.99\linewidth}
			\centerline{\includegraphics[width=\linewidth]{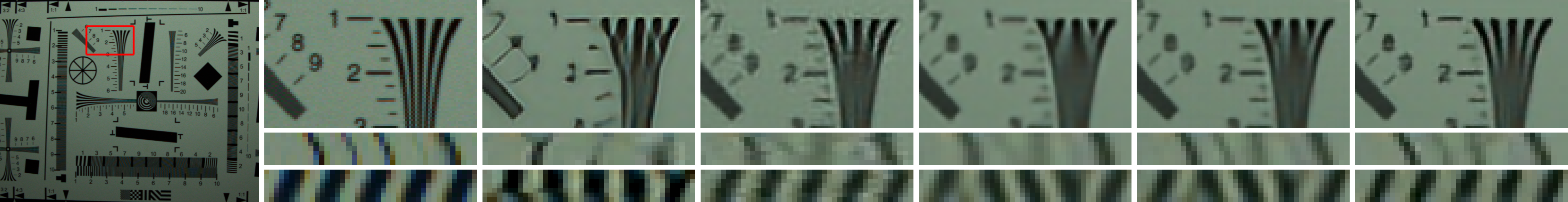}}
		\end{minipage}
		\begin{minipage}{.155\linewidth}
			\centerline{\textit{ISO Chart}}
		\end{minipage}
		\begin{minipage}{.13\linewidth}
			\centerline{PSNR/SSIM}
		\end{minipage}
		\begin{minipage}{.13\linewidth}
			\centerline{$27.92$/$0.853$}
		\end{minipage}
		\begin{minipage}{.13\linewidth}
			\centerline{$29.84$/$0.884$}
		\end{minipage}
		\begin{minipage}{.13\linewidth}
			\centerline{$30.61$/$0.901$}
		\end{minipage}
		\begin{minipage}{.13\linewidth}
			\centerline{$31.12$/$0.912$}
		\end{minipage}
		\begin{minipage}{.13\linewidth}
			\centerline{$31.74$/$0.919$}
		\end{minipage}

		\begin{minipage}{.99\linewidth}
			\centerline{\includegraphics[width=\linewidth]{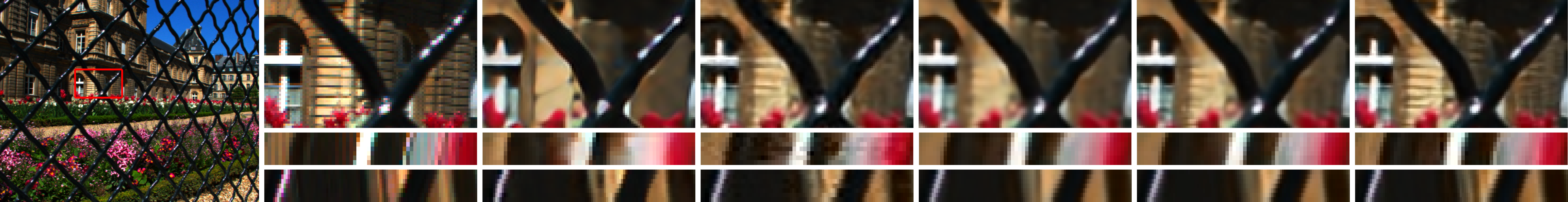}}
		\end{minipage}
		\begin{minipage}{.155\linewidth}
			\centerline{\textit{Palais}}
		\end{minipage}
		\begin{minipage}{.13\linewidth}
			\centerline{PSNR/SSIM}
		\end{minipage}
		\begin{minipage}{.13\linewidth}
			\centerline{$23.11$/$0.741$}
		\end{minipage}
		\begin{minipage}{.13\linewidth}
			\centerline{$24.72$/$0.802$}
		\end{minipage}
		\begin{minipage}{.13\linewidth}
			\centerline{$25.07$/$0.788$}
		\end{minipage}
		\begin{minipage}{.13\linewidth}
			\centerline{$25.57$/$0.811$}
		\end{minipage}
		\begin{minipage}{.13\linewidth}
			\centerline{$26.33$/$0.866$}
		\end{minipage}

		\caption{Comparison of $\times 4$ SR for real-world LF images, where the reconstructed central SAI and EPIs are shown. The average PSNR and SSIM of all views in one LF are also illustrated. For the real-world LF images, which have much noise, our model is able to recover more details and correct textures and also keep continuous lines in EPIs. By contrast, other methods show ambiguous results with overlapped details in both SAIs and EPIs.}
		\label{fig:genreal}
	\end{figure*}

\begin{figure*}[!t]
		\centering
		\scriptsize
		\begin{minipage}{.155\linewidth}
			\centerline{}
		\end{minipage}
		\begin{minipage}{.13\linewidth}
			\centerline{GT}
		\end{minipage}
		\begin{minipage}{.13\linewidth}
			\centerline{EDSR} 
		\end{minipage}
		\begin{minipage}{.13\linewidth}
			\centerline{resLF}
		\end{minipage}
		\begin{minipage}{.13\linewidth}
			\centerline{SAS}
		\end{minipage}
		\begin{minipage}{.13\linewidth}
			\centerline{4D}
		\end{minipage}
		\begin{minipage}{.13\linewidth}
			\centerline{Ours}
		\end{minipage}
		\begin{minipage}{.99\linewidth}
			\centerline{\includegraphics[width=\linewidth]{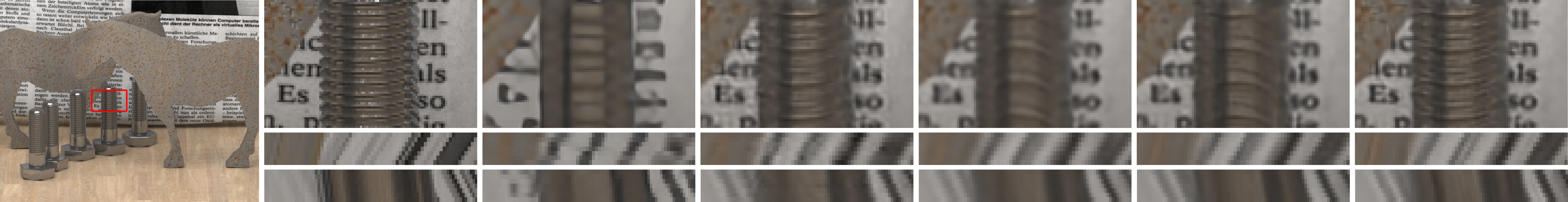}}
		\end{minipage}
		\begin{minipage}{.155\linewidth}
			\centerline{\textit{Horses}}
		\end{minipage}
		\begin{minipage}{.13\linewidth}
			\centerline{PSNR/SSIM}
		\end{minipage}
		\begin{minipage}{.13\linewidth}
			\centerline{$24.83$/$0.753$}
		\end{minipage}
		\begin{minipage}{.13\linewidth}
			\centerline{$27.07$/$0.812$}
		\end{minipage}
		\begin{minipage}{.13\linewidth}
			\centerline{$26.96$/$0.805$}
		\end{minipage}
		\begin{minipage}{.13\linewidth}
			\centerline{$27.59$/$0.829$}
		\end{minipage}
		\begin{minipage}{.13\linewidth}
			\centerline{$28.85$/$0.863$}
		\end{minipage}

		\begin{minipage}{.99\linewidth}
			\centerline{\includegraphics[width=\linewidth]{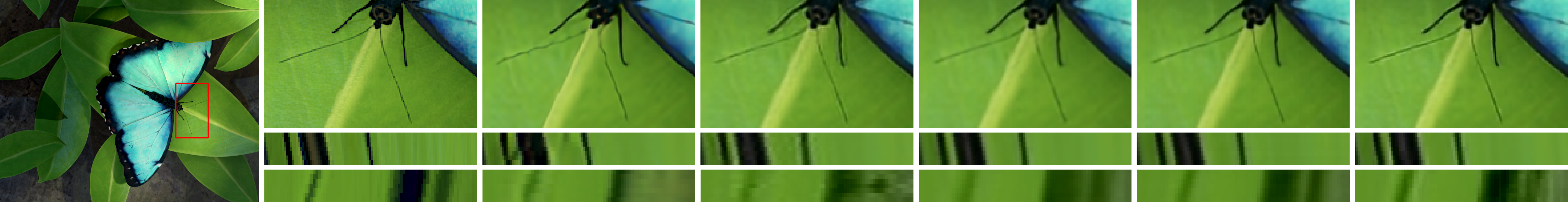}}
		\end{minipage}
		\begin{minipage}{.155\linewidth}
			\centerline{\textit{Papillon}}
		\end{minipage}
		\begin{minipage}{.13\linewidth}
			\centerline{PSNR/SSIM}
		\end{minipage}
		\begin{minipage}{.13\linewidth}
			\centerline{$35.70$/$0.938$}
		\end{minipage}
		\begin{minipage}{.13\linewidth}
			\centerline{$38.05$/$0.959$}
		\end{minipage}
		\begin{minipage}{.13\linewidth}
			\centerline{$38.71$/$0.960$}
		\end{minipage}
		\begin{minipage}{.13\linewidth}
			\centerline{$38.52$/$0.965$}
		\end{minipage}
		\begin{minipage}{.13\linewidth}
			\centerline{$39.83$/$0.969$}
		\end{minipage}

		\begin{minipage}{.99\linewidth}
			\centerline{\includegraphics[width=\linewidth]{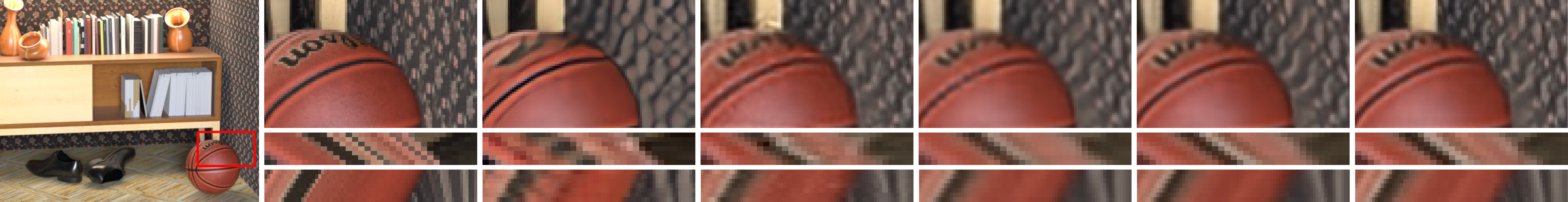}}
		\end{minipage}
		\begin{minipage}{.155\linewidth}
			\centerline{\textit{Sideboard}}
		\end{minipage}
		\begin{minipage}{.13\linewidth}
			\centerline{PSNR/SSIM}
		\end{minipage}
		\begin{minipage}{.13\linewidth}
			\centerline{$24.53$/$0.731$}
		\end{minipage}
		\begin{minipage}{.13\linewidth}
			\centerline{$27.54$/$0.842$}
		\end{minipage}
		\begin{minipage}{.13\linewidth}
			\centerline{$27.43$/$0.813$}
		\end{minipage}
		\begin{minipage}{.13\linewidth}
			\centerline{$28.10$/$0.845$}
		\end{minipage}
		\begin{minipage}{.13\linewidth}
			\centerline{$29.06$/$0.884$}
		\end{minipage}
		
		\caption{Comparison of $\times 4$  SR for synthetic LF images, where central SAI and EPIs are shown. Our model can recover the correct high-frequency details. In the synthetic LF images, we reconstruct sharp super-resolution picture, while other models can not recover the high-frequency details, or even recover the texture in wrong directions. The average PSNR and SSIM of all SAIs in one LF are also illustrated. }
		
		\label{fig:synthetic}
	\end{figure*}

	We compare our method with other state-of-the-art methods for $\times 2$ and $\times 4$ tasks on different LF datasets.
	Quantitative comparisons are illustruted in Table.~\ref{x2compare} and Table.~\ref{x4compare}. 
	Qualitative results are shown in Fig.~\ref{fig:genreal} and Fig.~\ref{fig:synthetic}, in which the EPIs from the reconstructed LF are also shown. The single image SR method EDSR~\cite{lim2017enhanced} is implemented on each SAI separately. Since other sub-pixel information in other SAIs is not combined, it cannot achieve comparable results with other LFSSR methods. 
	In resLF~\cite{zhang2019residual}, each SAI is super-resolved separately using different models, so that the EPIs show discontinuous lines, which means the consistency between each SAI is hard to preserve.
	Moreover, since only parts of SAIs are used as input, it cannot extract enough information for LFSSR.
	Using 4D convolution, the network~\cite{yeung2018light} achieves higher results than their SAS convolution, which indicates that using 2D convolution iteratively on the angular domain and spatial domain cannot effectively extract LF features. 
	Moreover, since prior is not provided to 4D convolution insufficiently, the model~\cite{yeung2018light} cannot distinguish features from different dimensions and fuse them effectively.
	In LFNet~\cite{wang2018lfnet}, only SAIs in central horizontal and vertical SAI stacks are used to extract features. 
	The features of these SAIs are supposed to have different offsets for other SAIs recovery in the stacked generalization. 
	Therefore, reconstruction qualities are relatively low.
	
	\begin{table}[!h]
		\centering
		\caption{Average PSNR/SSIM in different datasets for $\times 4$ LFSSR.}
		\begin{tabular}{c|c|c|c|c|c}
			\hline
			Methods  & EDSR~\cite{lim2017enhanced} & SAS~\cite{yeung2018light} & 4D~\cite{yeung2018light} & resLF~\cite{zhang2019residual} & Ours \\
			\hline
			\hline
			\textit{HCI1} & 31.95/0.866 & 34.73/0.906 & 34.82/0.918 & 34.43/0.909 & \textbf{36.02/0.932} \\
			\hline
			\textit{HCI2} & 28.03/0.783 & 30.59/0.841 & 30.99/0.862 & 30.31/0.853 & \textbf{31.57/0.886} \\
			\hline
			\textit{EPEL}  & 27.77/0.791 & 29.95/0.838 & 30.19/0.851 & 29.28/0.833 & \textbf{30.62/0.875} \\
			\hline
			\textit{General} & 31.33/0.869 & 34.78/0.918 & 35.08/0.926 & 33.86/0.909 & \textbf{36.01/0.940} \\
			\hline
		\end{tabular}%
		\label{x4compare}%
	\end{table}%

	By comparison, our method achieves the highest PSNR and SSIM results in LFSSR. 
	In real-world dataset \textit{General}, the PSNR values of MDFN are around $1$ dB higher in both $\times 2$ and $\times 4$ LFSSR than 4D~\cite{yeung2018light}, which is the current best module in LFSSR. 
	Moreover, although our model is trained with real-world images, it is still effective for synthetic images in \textit{HCI1} and \textit{HCI2}.
	In Fig.~\ref{fig:genreal} and Fig.~\ref{fig:synthetic}, our proposed MDFN also achieves state-of-the-art performance compared to the other deep-learning-based algorithms in both synthetical and real-world dataset. 
	In our reconstructed SAIs, details are accurately recovered without aliasing.
	Our EPIs also show distinct lines, which means that our model is able to keep the instinct structure of LFs and keep accurate disparity information.

	\begin{table}[!h]
		\centering
		
		\caption{Parameters number of different methods for $\times 2$ LFSSR.}
		
		\begin{tabular}{c|c|c|c}
			\hline
			Methods & 4D~\cite{yeung2018light} & SAS~\cite{yeung2018light}  & Ours\\
			\hline
			Parameters & 3,355,920 & 776,640 & \textbf{498,466}\\
			\hline
			PSNR/SSIM& 42.09/0.982 & 41.87/0.982 & \textbf{43.01}/\textbf{0.986}\\
			\hline
		\end{tabular}
		
		\label{parameterscompare}
	\end{table}

	% Moreover, the parameters of our network are less than  4D's. 
	
	We also compare the number of parameters for different models in Table.~\ref{parameterscompare}. 
	As shown, our network has significantly fewer parameters, \textit{e.g.} a sixth of 4D~\cite{yeung2018light},  but achieves better performance than other state-of-art modules.
	%========================================================================

	\section{Conclusions}
	
	In this paper, we introduce a novel learning-based framework for LFSSR to extract features from multi-dimension in parallel and effectively fused them together to offer high-frequency information.
	Different from traditional upsampling methods, we also propose dynamic upsampling filters to mix values from different SAIs according to their specific features, in which the disparity information is implicitly considered.
	The high-frequency residual information is also produced and combined with the upsampled LFs to output the final full LFs with high spatial resolution.
	In the experiments, we show the importance of parallelization in the proposed MDFA to extract multi-dimension features. The proposed dynamic filters are also visualized and verified that it performs better than other upsample methods. 
	Experimental results demonstrate that the proposed method outperforms the state-of-the-art methods by a large margin in various challenging natural scenes, which is able to recover LF images with sharp details and also maintains the consistency of EPIs.

	\clearpage
	% ---- Bibliography ----
	%
	% BibTeX users should specify bibliography style 'splncs04'.
	% References will then be sorted and formatted in the correct style.
	%
	\bibliographystyle{splncs04}
	\bibliography{egbib}

\begin{thebibliography}{10}
\providecommand{\url}[1]{\texttt{#1}}
\providecommand{\urlprefix}{URL }
\providecommand{\doi}[1]{https://doi.org/#1}

\bibitem{bishop2009light}
Bishop, T.E., Zanetti, S., Favaro, P.: Light field superresolution. In: 2009
  IEEE International Conference on Computational Photography (ICCP). pp.~1--9.
  IEEE (2009)

\bibitem{caballero2017real}
Caballero, J., Ledig, C., Aitken, A., Acosta, A., Totz, J., Wang, Z., Shi, W.:
  Real-time video super-resolution with spatio-temporal networks and motion
  compensation. In: Proceedings of the IEEE Conference on Computer Vision and
  Pattern Recognition. pp. 4778--4787 (2017)

\bibitem{cheng2019light}
Cheng, Z., Xiong, Z., Liu, D.: Light field super-resolution by jointly
  exploiting internal and external similarities. IEEE Transactions on Circuits
  and Systems for Video Technology  (2019)

\bibitem{cho2013modeling}
Cho, D., Lee, M., Kim, S., Tai, Y.W.: Modeling the calibration pipeline of the
  lytro camera for high quality light-field image reconstruction. In:
  Proceedings of the IEEE International Conference on Computer Vision. pp.
  3280--3287 (2013)

\bibitem{farrugia2019light}
Farrugia, R., Guillemot, C.: Light field super-resolution using a low-rank
  prior and deep convolutional neural networks. IEEE transactions on pattern
  analysis and machine intelligence (TPAMI)  (2019)

\bibitem{he2015delving}
He, K., Zhang, X., Ren, S., Sun, J.: Delving deep into rectifiers: Surpassing
  human-level performance on imagenet classification. In: Proceedings of the
  IEEE international conference on computer vision. pp. 1026--1034 (2015)

\bibitem{honauer2016benchmark}
Honauer, K., Johannsen, O., Kondermann, D., Goldluecke, B.: A dataset and
  evaluation methodology for depth estimation on light fields. In: Proceedings
  of the Asian Conference on Computer Vision (ACCV). pp. 19--34. Springer
  (2016)

\bibitem{jo2018deep}
Jo, Y., Wug~Oh, S., Kang, J., Joo~Kim, S.: Deep video super-resolution network
  using dynamic upsampling filters without explicit motion compensation. In:
  Proceedings of the IEEE conference on computer vision and pattern
  recognition. pp. 3224--3232 (2018)

\bibitem{johannsen2017taxonomy}
Johannsen, O., Honauer, K., Goldluecke, B., Alperovich, A., Battisti, F., Bok,
  Y., Brizzi, M., Carli, M., Choe, G., Diebold, M., et~al.: A taxonomy and
  evaluation of dense light field depth estimation algorithms. In: Proceedings
  of the IEEE Conference on Computer Vision and Pattern Recognition Workshops
  (CVPRW). pp. 82--99 (2017)

\bibitem{kim2016accurate}
Kim, J., Jung~Kwon, L., Kyoung~Mu, L.: Accurate image super-resolution using
  very deep convolutional networks. In: Proceedings of the IEEE Conference on
  Computer Vision and Pattern Recognition (CVPR). pp. 1646--1654 (2016)

\bibitem{kim2018spatio}
Kim, T.H., Sajjadi, M.S., Hirsch, M., Sch{\"o}lkopf, B.: Spatio-temporal
  transformer network for video restoration. In: European Conference on
  Computer Vision. pp. 111--127. Springer (2018)

\bibitem{lim2017enhanced}
Lim, B., Son, S., Kim, H., Nah, S., Mu~Lee, K.: Enhanced deep residual networks
  for single image super-resolution. In: Proceedings of the IEEE conference on
  computer vision and pattern recognition workshops (CVPRW). pp. 136--144
  (2017)

\bibitem{Lytro}
Ng, R.: Lytro redefines photography with light field cameras.
  \url{http://www.lytro.com}, accessed: Oct. 22, 2018

\bibitem{ng2005light}
Ng, R., Levoy, M., Br{\'e}dif, M., Duval, G., Horowitz, M., Hanrahan, P.,
  et~al.: Light field photography with a hand-held plenoptic camera. Computer
  Science Technical Report CSTR  \textbf{2}(11),  1--11 (2005)

\bibitem{Raytrix}
Perwaß, C., Wietzke, L.: Raytrix: Light filed technology.
  \url{http://www.raytrix.de}, accessed: Oct. 22, 2018

\bibitem{rerabek2016new}
Rerabek, M., Ebrahimi, T.: New light field image dataset. In: 8th International
  Conference on Quality of Multimedia Experience (2016)

\bibitem{rossi2018geometry}
Rossi, M., Frossard, P.: Geometry-consistent light field super-resolution via
  graph-based regularization. IEEE Transactions on Image Processing
  \textbf{27}(9),  4207--4218 (2018)

\bibitem{Stanford2016}
Sunder~Raj, A., Lowney, M., Shah, R., Wetzstein, G.: The stanford lytro light
  field archive. \url{http://lightfields.stanford.edu/LF2016.html} (2016),
  accessed: Oct. 22, 2018

\bibitem{wang2019edvr}
Wang, X., Chan, K.C., Yu, K., Dong, C., Change~Loy, C.: Edvr: Video restoration
  with enhanced deformable convolutional networks. In: Proceedings of the IEEE
  Conference on Computer Vision and Pattern Recognition Workshops. pp.~0--0
  (2019)

\bibitem{wang2018lfnet}
Wang, Y., Liu, F., Zhang, K., Hou, G., Sun, Z., Tan, T.: Lfnet: A novel
  bidirectional recurrent convolutional neural network for light-field image
  super-resolution. IEEE Transactions on Image Processing  \textbf{27}(9),
  4274--4286 (2018)

\bibitem{wang2004image}
Wang, Z., Bovik, A.C., Sheikh, H.R., Simoncelli, E.P.: Image quality
  assessment: from error visibility to structural similarity. IEEE Transactions
  on Image Processing  \textbf{13}(4),  600--612 (2004)

\bibitem{wanner2012spatial}
Wanner, S., Goldluecke, B.: Spatial and angular variational super-resolution of
  4d light fields. In: Proceedings of the European Conference on Computer
  Vision (ECCV). pp. 608--621. Springer (2012)

\bibitem{wanner2014variational}
Wanner, S., Goldluecke, B.: Variational light field analysis for disparity
  estimation and super-resolution. IEEE Transactions on Pattern Analysis and
  Machine Intelligence (TPAMI)  \textbf{36}(3),  606--619 (2014)

\bibitem{wanner2013datasets}
Wanner, S., Meister, S., Goldluecke, B.: Datasets and benchmarks for densely
  sampled 4d light fields. In: Vision, Modeling \& Visualization. vol.~13, pp.
  225--226. Citeseer (2013)

\bibitem{xue2019video}
Xue, T., Chen, B., Wu, J., Wei, D., Freeman, W.T.: Video enhancement with
  task-oriented flow. International Journal of Computer Vision
  \textbf{127}(8),  1106--1125 (2019)

\bibitem{yeung2018light}
Yeung, H.W.F., Hou, J., Chen, X., Chen, J., Chen, Z., Chung, Y.Y.: Light field
  spatial super-resolution using deep efficient spatial-angular separable
  convolution. IEEE Transactions on Image Processing  \textbf{28}(5),
  2319--2330 (2018)

\bibitem{yoon2017light}
Yoon, Y., Jeon, H.G., Yoo, D., Lee, J.Y., Kweon, I.S.: Light-field image
  super-resolution using convolutional neural network. IEEE Signal Processing
  Letters  \textbf{24}(6),  848--852 (2017)

\bibitem{plex_vr}
Yu, J., Hong, X., Yang, J., Ma, Y.: Dgene: The light of science, the light of
  future. \url{http://www.plex-vr.com/product/model/}, accessed: Oct. 22, 2018

\bibitem{zhang2019residual}
Zhang, S., Lin, Y., Sheng, H.: Residual networks for light field image
  super-resolution. In: Proceedings of the IEEE Conference on Computer Vision
  and Pattern Recognition (CVPR). pp. 11046--11055 (2019)

\bibitem{zhang2016robust}
Zhang, S., Sheng, H., Li, C., Zhang, J., Xiong, Z.: Robust depth estimation for
  light field via spinning parallelogram operator. Computer Vision and Image
  Understanding  \textbf{145},  148--159 (2016)

\bibitem{zhang2018micro}
Zhang, S., Sheng, H., Yang, D., Zhang, J., Xiong, Z.: Micro-lens-based matching
  for scene recovery in lenslet cameras. IEEE Transactions on Image Processing
  \textbf{27}(3),  1060--1075 (2017)

\end{thebibliography}
\end{document}